\documentstyle[aps,preprint,pra]{revtex}
\draft
\input{epsf}
\tighten
\begin{document}
\title{Statistical uncertainty in quantum optical photodetection
measurements}
\author{Konrad Banaszek}
\address{Instytut Fizyki Teoretycznej, Uniwersytet Warszawski,
Ho\.{z}a 69, PL-00-681 Warszawa, Poland}
\date{\today}
\maketitle
\begin{abstract}
We present a complete statistical analysis of quantum optical
measurement schemes based on photodetection. Statistical distributions
of quantum observables determined from a finite number of experimental
runs are characterized with the help of the generating function, which
we derive using the exact statistical description of 
raw experimental outcomes. We use the
developed formalism to point out that the statistical uncertainty results
in substantial limitations of the determined information on the
quantum state: though a family of observables characterizing the
quantum state can be safely
evaluated from experimental data, its further use to obtain the
expectation value of some operators generates exploding statistical
errors. These issues are discussed using the example of
phase--insensitive measurements of a single light mode.
We study reconstruction of the photon
number distribution from photon counting and random phase homodyne
detection. We show that utilization of the reconstructed distribution
to evaluate a simple well--behaved observable, namely the parity
operator, encounters difficulties due to accumulation of statistical
errors.  As the parity operator yields the Wigner function at the phase
space origin, this example also demonstrates that transformation
between various experimentally determined representations of the
quantum state is a quite delicate matter.
\end{abstract}

\section{Introduction}

Over recent years, the set of tools for measuring quantum statistical
properties of optical radiation has substantially enlarged. The
experimental demonstration of optical homodyne tomography
\cite{SmitBeckPRL93} has been followed by detailed studies of this
beautiful technique 
\cite{OregonOHT,KonstanzOHT,PatternFunctions,LeonMunrOC96,%
DAriMaccQSO97,DAriPariPLA97,LeonMunrPRA96}, and diverse
novel schemes have been proposed
\cite{Endoscopy,PaulTormPRL96,WallVogePRA96,BanaWodkPRL96,%
ZuccVogePRA96,JacoKnigJMO97}.
The quantum optical ``toolbox'' for measuring light contains now
experimentally established schemes for reconstructing various
representations of its quantum state: the $Q$ function
\cite{DblHomExp}, the Wigner function, and the density matrix in the
quadrature and the Fock bases 
\cite{SmitBeckPRL93,OregonOHT,KonstanzOHT}.  A
device that is used in most of quantum optical schemes to convert the
quantum signal to a macroscopic level is the photodetector. Thus,
photodetection is a basic ingredient of quantum optical measurements.

The quantum state can be characterized using various representations:
quasidistribution functions or a density matrix in a specific basis. 
From a theoretical point of view, all these forms are equivalent. Each
representation contains complete information on the quantum state, and
they can be transformed from one to another. Any observable related to
the measured systems can be evaluated from an arbitrary representation
using an appropriate expression.

This simple picture becomes much more complicated when we deal with
real experimental data rather than analytical formulae. Each quantity
determined from a finite number of experimental runs is affected
by a statistical error. Consequently, the density matrix or the
quasidistribution function reconstructed from the experimental data
is known only with some statistical uncertainty. This uncertainty
is important when we further use the reconstructed information to
calculate other observables or to pass to another representation.
The crucial question is, whether determination of a certain representation
with sufficient accuracy guarantees that arbitrary observable can be
calculated from these data with a reasonably low statistical error. If this
is not the case, the reconstructed information on the quantum state of
the measured system turns out to be somewhat incomplete. Furthermore,
transformation between various representations of the quantum state
becomes a delicate matter.

These and related problems call for a rigorous statistical analysis of
quantum optical measurements. The purpose of this paper is to provide a
complete statistical description of a measurement of quantum observables
in optical schemes based on photodetection. This approach
fully characterizes statistical properties of quantities determined
in a realistic measurement from a finite number of experimental
runs. It can be applied either to determination of a single quantum
observable, or to the reconstruction of the quantum state in a specific
representation. Within the presented framework we study, using  a simple
example, the completeness of the experimentally reconstructed information
on the quantum state. We demonstrate the pathological behavior suggested
above, when the reconstructed data cannot be used to calculate some
observables due to rapidly exploding statistical errors.

An important issue is the statistical methodology applied to retrieve
information on the measured quantum system from experimental data. Various
approaches have been recently developed, based on the maximum entropy
principle \cite{MaxEntropy}, 
the least--squares inversion \cite{LeastSquares}, 
and the maximum likelihood
estimation \cite{MaxLikelihood,BanaszekMaxLik}.
In this paper we will consider the most straightforward and
so far the most commonly used strategy, where experimental relative
frequencies are taken as
estimates for quantum mechanical probability
distributions. 
Quantum observables are reconstructed from these data via linear
transformations.

Selected aspects of statistical uncertainty resulting from the finite
character of the experimental data sample have been discussed in some
particular cases. Systematic and statistical errors in homodyne
measurements of the density matrix were studied 
\cite{LeonMunrOC96,DAriMaccQSO97},
and it was found that homodyne detection of quantum observables is
accompanied by excess noise compared to a direct measurement
\cite{DAriPariPLA97}. Analysis of a photon counting scheme for
sampling quantum phase space showed that compensation for the nonunit
detection efficiency is in general not possible \cite{BanaWodkJMO97}.
The present paper provides a general statistical analysis of quantum
optical schemes based on photodetection.

This paper is organized
as follows. The starting point of our analysis is the probability
distribution of obtaining a specific histogram from $N$ runs of the
experimental setup. This basic quantity determines all statistical
properties of quantum observables reconstructed from a finite sample of
experimental data. We characterize these properties using the
generating function, for which we derive an exact expression
directly from the probability distribution of the experimental
outcomes. These general results are presented in Sec.~\ref{Sec:Analysis}.
Then, in Sec.~\ref{Sec:Random}, we use the developed formalism to discuss
the reconstruction of the photon number distribution 
of a single light mode, and its subsequent utilization to
evaluate the parity operator $\hat{\Pi}$. We consider two experimental
schemes: direct photon counting using an imperfect detector, and
homodyne detection with random phase. In both the cases we
find that the evaluation of the parity operator from the reconstructed
photon statistics is a very delicate matter. For photon counting of a
thermal state, we show that neither the statistical mean value 
of $\hat{\Pi}$ nor its
variance have to exist when we take into account arbitrarily high count
numbers.  For random phase homodyne detection, the statistical error of
the parity operator is an interplay of the number of runs $N$ and the
specific regularization method used for its evaluation. The example
of the parity operator illustrates
difficulties related to the transformations
between various experimentally determined representations of the
quantum state, as the parity operator yields, up to a multiplicative
constant, the Wigner function at the phase space origin. Finally,
Sec.~\ref{Sec:Conclusions} summarizes the paper.

\section{Statistical analysis of experiment}
\label{Sec:Analysis}

In photodetection measurements,
the raw quantity delivered by a single experimental run is the number of
photoelectrons ejected from the active material of the detectors. The
data recorded for further processing depends on a specific scheme. It
may be just the number of counts on a single detector, or a
difference of photocounts on a pair of photodetectors, which
is the case of balanced homodyne detection. It may also be a finite
sequence of integer numbers, e.g.\ for double homodyne detection. We
will denote in general this data by $n$, keeping in mind all the
possibilities.

The experimental scheme may have some external parameters $\theta$,
for example the phase of the local oscillator in homodyne detection.
The series of measurements are repeated for various settings
$\theta_i$ of these parameters. Thus, what is eventually obtained from the
experiment, is a set of histograms $\{k_n\}_{\theta_i}$, telling in how
many runs with the settings $\theta_i$ the outcome $n$ has been recorded.
We assume that for each setting the same total number of $N$ runs has
been performed.

The theoretical probability $p_n(\theta_i)$ of
obtaining the outcome $n$ in a run with settings $\theta_i$ is given by
the expectation value of a positive operator valued measure (POVM)
$\hat{p}_{n}(\theta_i)$ acting in the Hilbert space of the measured
system. The reconstruction of an observable $\hat{A}$ is possible, if it
can be represented as a linear combination of the POVMs for the
settings used in the experiment:
\begin{equation}
\label{Eq:hatAdef} 
\hat{A} = \sum_{i} \sum_{n}
a_{n}(\theta_{i})\hat{p}_{n}(\theta_i),
\end{equation} 
where $a_{n}(\theta_i)$ are the kernel functions. 
The above formula allows
one to compute the quantum expectation value $\langle\hat{A}\rangle$
from the probability distributions $p_n(\theta_i)$.

This theoretical relation has to be applied now to the experimental
data. The simplest and the most commonly used strategy is to estimate
the probability distributions $p_n(\theta_i)$
by experimental relative frequencies
$(k_n/N)_{\theta_i}$. The relative frequencies integrated with the
appropriate kernel functions yield an estimate for the expectation
value of the operator $\hat{A}$. For simplicity, we will denote this
estimate just by $A$. Thus, the recipe for reconstructing the observable
$A$ from experimental data is given by 
the counterpart of Eq.~(\ref{Eq:hatAdef}):
\begin{equation}
\label{Eq:Adeterm}
A = \sum_{i} \sum_{n} a_{n}(\theta_{i}) \left(\frac{k_n}{N}
\right)_{\theta_i}.
\end{equation}

We will now analyse statistical properties of the
observable $A$ evaluated according to Eq.~(\ref{Eq:Adeterm})
from data collected in a finite number of experimental runs. 
Our goal is to characterize the statistical distribution $w(A)$
defining the probability that the experiment yields a specific
result $A$.
The fundamental object in this
analysis is the probability ${\cal P}(\{k_n\};\theta)$ of
obtaining a specific histogram $\{k_n\}$ for the settings $\theta$. 
In order to avoid convergence problems, we will
restrict the possible values of $n$ to a finite set by introducing
a cut--off. The probability
${\cal P}(\{k_n\};\theta)$ is then given by the multinomial distribution:
\begin{equation}
{\cal P}(\{k_n\};\theta)
=
\frac{N!}{(N-\sum_{n}'k_n)!}
\left( 1 - {\sum_{n}}'p_n(\theta) \right)^{N-\sum_{n}' k_n}
{\prod_{n}}'\frac{1}{k_n!} [p_n(\theta)]^{k_n},
\end{equation}
where prim in sums and products denotes the cut--off.

Let us first consider a contribution $A_i$ to the observable
$A$ calculated from the histogram $\theta_i$:
\begin{equation}
A_i = {\sum_{n}}' a_n (\theta_i) \left( \frac{k_n}{N} 
\right)_{\theta_i}.
\end{equation}
Its statistical distribution $w(A_i; \theta_i)$ is given by the 
following sum over all possible histograms that can be
obtained from $N$ experimental runs:
\begin{equation}
w(A_i;\theta_i) = \sum_{\{k_n\}} {\cal P}(\{k_n\}; \theta_i)
\delta\left( A_i - \frac{1}{N} {\sum_{n}}'a_n(\theta_i) k_n \right).
\end{equation}
Equivalently, the statistics of $A_i$ can be characterized by
the generating function $\tilde{w}(\lambda;\theta_i)$ for the moments,
which is the Fourier transform of the distribution $w(A_i;\theta_i)$:
\begin{eqnarray}
\tilde{w}(\lambda;\theta_i) & =  & 
\int \text{d}A_i \; e^{i\lambda A_i}
w(A_i;\theta_i)
\nonumber \\
& = & \sum_{\{k_n\}} {\cal P}(\{k_n\}; \theta_i)
\exp \left( \frac{i\lambda}{N}
{\sum_{n}}' a_n(\theta_i) k_n \right). 
\end{eqnarray}
An easy calculation yields the explicit form of the generating function:
\begin{equation}
\tilde{w}(\lambda; \theta_i) = \left(
1 + {\sum_{n}}' p_n(\theta_i)(e^{i\lambda a_n(\theta_i)/N}-1)
\right)^{N}.
\end{equation}
The observable $A$ is obtained via summation of the components $A_i$
corresponding to all settings of the external parameters $\theta_i$. As these
components are determined from disjoint subsets of the experimental
data, they are statistically independent. Consequently, the generating
function $\tilde{w}(\lambda)$ for the moments of the observable $A$
is given by the product:
\begin{eqnarray}
\tilde{w}(\lambda) & = & 
\int\text{d}A \; e^{i\lambda A} w(A) =
\prod_{i} \tilde{w}(\lambda; \theta_i)
\nonumber \\
& = &
\prod_{i}
\left(
1 + {\sum_{n}}' p_n(\theta_i)(e^{i\lambda a_n(\theta_i)/N}-1)
\right)^{N}.
\end{eqnarray}
This expression contains the complete statistical information
on determination of the observable $A$ from a finite number
of runs of a specific experimental setup. The measuring apparatus is
included in this expression in the form of a family of POVMs
$\hat{p}_{n}(\theta_i)$. The quantum expectation value of these POVMs
over the state of the measured system yields the probability
distributions $p_n(\theta_i)$. Finally, the coefficients
$a_n(\theta_i)$ are given by the computational recipe for reconstructing
the observable $A$ from the measured distributions.

The basic characteristics of statistical properties of the observable $A$ is
provided by the mean value $\text{E}(A)$ 
and the variance $\text{Var}(A)$. These
two quantities can be easily found by differentiating the logarithm of the
generating operator:
\begin{eqnarray}
\text{E}(A) & = & \left. \frac{1}{i} \frac{\text{d}}{\text{d}\lambda}
\log \tilde{w}(\lambda) \right|_{\lambda=0} =
\sum_{i} {\sum_{n}}' a_n(\theta_i) p_{n}(\theta_i), \\
\text{Var}(A) & = & \left.
\frac{1}{i^2} \frac{\text{d}^2}{\text{d}\lambda^2}
\log\tilde{w}(\lambda) \right|_{\lambda=0} \nonumber \\
& = & 
\label{Eq:VarA}
\frac{1}{N}
\left[ \sum_{i}
{\sum_{n}}' a_n^{2} (\theta_i) p_{n}(\theta_i)
- 
\sum_{i}
\left( {\sum_{n}}' a_n(\theta_i) p_n(\theta_i) \right)^{2}
\right].
\end{eqnarray}
The statistical error is scaled with the inverse of the square root
of the number of runs $N$. Let us note that the second component in the
derived formula for $\text{Var}(A)$ differs from that used in 
the discussions of homodyne tomography in
Refs.~\cite{DAriMaccQSO97,DAriPariPLA97},
where it was equal just to $[\text{E}(A)]^2$.
This difference results from different assumptions about the local
oscillator phase: in Refs.~\cite{DAriMaccQSO97,DAriPariPLA97} 
it was considered to be a stochastic variable in order to avoid systematic
errors, whereas we have assumed that
the number of runs is fixed for each selected setting of the
external parameters.

The goal of quantum state measurements is to retrieve the maximum amount of
information on the quantum state available from the experimental data.
Therefore the experimental histograms are usually processed many times in
order to reconstruct a family of observables characterizing the quantum
state. Of course, quantities determined from the same set of 
experimental data are not
statistically independent, but in general may exhibit correlations. 
The analysis presented above can be easily extended to evaluation of any
number of observables from the experimental data. If we restrict our
attention to the basic, second--order characterization of these correlations,
it is sufficient to discuss determination of two observables. 
Let us suppose that
in addition to $A$, another observable $B$ has been calculated
from the histograms $\{k_n\}_{\theta_i}$ according to the formula:
\begin{equation}
B = \sum_{i} {\sum_{n}}' b_{n} (\theta_i) \left( \frac{k_n}{N}
\right)_{\theta_i}.
\end{equation}
The generating function $\tilde{w}(\lambda,\mu)$ corresponding
to the joint probability distribution $w(A,B)$ can be found analogously
to the calculations presented above. The final result is:
\begin{eqnarray}
\tilde{w}(\lambda,\mu) & = & \int \text{d}A \text{d}B \;
e^{i\lambda A + i \mu B} w(A,B) \nonumber \\
& = &  
\prod_{i} \left( 1 + {\sum_{n}}' p_n(\theta_i)
(e^{i\lambda a_n(\theta_i)/N + i\mu b_n(\theta_i)/N} - 1 )
\right)^{N}.
\end{eqnarray}
The covariance between the experimentally determined values
of $A$ and $B$ is given by:
\begin{eqnarray}
\text{Cov}(A,B) & = &
\left.
\frac{1}{i^2} \frac{\text{d}^2}{\text{d}\lambda
\text{d}\mu} \log \tilde{w}(\lambda,\mu)
\right|_{\lambda,\mu=0} \nonumber \\
& = & 
\frac{1}{N}
\sum_{i} \left[ {\sum_{n}}' a_n (\theta_i) b_n (\theta_i)
p_n (\theta_i) - \left(
{\sum_{n}}' a_n(\theta_i) p_n(\theta_i) \right)
\left({\sum_{m}}' b_{m}(\theta_i) p_{m}(\theta_i) \right)
\right]. \nonumber \\
& &
\end{eqnarray}
The covariance can be normalized to the interval $[-1,1]$ using
$\text{Var}(A)$ and $\text{Var}(B)$, which yields the correlation
coefficient for the pair of observables $A$ and $B$:
\begin{equation}
\label{Eq:Corr}
\text{Corr}(A,B) = \frac{\text{Cov}(A,B)}{\sqrt{\text{Var}(A)
\text{Var}(B)}}
\end{equation}
This quantity defines whether the statistical deviations of $A$ and
$B$ tend to have the same or opposite sign, which corresponds
respectively to the positive or negative value of $\text{Corr}(A,B)$. 

We have assumed that the histograms $k_n$ have been measured for
a finite number of external parameters settings $\theta_i$, which
is always the case in an experiment. However, in some schemes the
measurement of histograms is in principle necessary for all values of a
continuous parameter. For example, in optical homodyne tomography the full
information on the quantum state is contained in a family of quadrature
distributions for all local oscillator phases.  Restriction to a finite
set of phases introduces a systematic error to the measurement
\cite{DAriMaccQSO97,LeonMunrPRA96}.

\section{Phase--insensitive detection of a light mode}
\label{Sec:Random}

We will now apply the general formalism developed in the preceding
section to the reconstruction of phase--independent properties
of a single light
mode. The basic advantage of this exemplary system is that it
will allow us to discuss, in a very transparent way, pathologies
resulting from the statistical uncertainty.
All phase--independent properties of a single light mode are fully
characterized by its photon number distribution $\rho_\nu$.
Therefore, it is sufficient to apply a
phase--insensitive technique to measure the photon statistics of the field. 
We will consider two measurement schemes that can be used for this
purpose: direct photon counting and
random phase homodyne detection.

The photon number distribution is given by the expectation value of a
family of projection operators 
$\hat{\rho}_{\nu} = |\nu\rangle\langle\nu|$,
where $|\nu\rangle$ is the $\nu$th Fock state. In principle, 
knowledge of this
distribution enables us to evaluate any phase--independent
observable related to the measured field. A simple yet nontrivial
observable, which we will use to point out difficulties with the
completeness of the reconstructed information on the quantum state,
is the parity operator:
\begin{equation}
\label{Eq:ParityOp}
\hat{\Pi} = \sum_{\nu=0}^{\infty} (-1)^{\nu} |\nu\rangle\langle\nu|.
\end{equation}
This operator is bounded, and well defined on the complete Hilbert
space of a single light mode. Its expectation value is given by the
alternating series of the photon number distribution:
\begin{equation}
\label{Eq:Parity}
\langle\hat{\Pi}\rangle = \sum_{\nu=0}^{\infty} (-1)^{\nu} 
\langle\hat{\rho}_{\nu}\rangle,
\end{equation}
which is absolutely convergent for any quantum state. Therefore, any
pathologies connected to its determination
from experimental data, if there appear any, cannot be
ascribed to its singular analytical properties.

\subsection{Direct photon counting}

First, we will consider the reconstruction of phase insensitive
properties of a single light mode from data measured using a
realistic, imperfect photodetector. 
The positive operator--valued measure
$\hat{p}_n$ describing the probability of ejecting $n$ photoelectrons
from the detector is given by \cite{KellKleiPR64}:
\begin{equation}
\hat{p}_{n} = \; : \frac{(\eta\hat{a}^{\dagger}\hat{a})^{n}}{n!}
\exp(-\eta\hat{a}^{\dagger}\hat{a}):,
\end{equation}
where $\hat{a}$ is the annihilation operator of the light mode, and
$\eta$ is the quantum efficiency of the photodetector. 
In the limit $\eta\rightarrow 1$ we get directly $\hat{p}_{n}
= |n\rangle\langle n|$. In a general case, the probability
distribution for the photoelectron number 
is related to the photon statistics
via the Bernoulli transformation.
This relation can be analytically inverted
\cite{KissHerzPRA95},
which yields the expression:
\begin{equation}
\label{Eq:InvBernoulli}
\hat{\rho}_{\nu} = \sum_{n=0}^{\infty} r_{\nu n}^{(\eta)} \hat{p}_{n},
\end{equation}
where the kernel functions $r_{\nu n}^{(\eta)}$
are given by:
\begin{equation}
r_{\nu n}^{(\eta)} = \left\{
\begin{array}{ll}
0, & n < \nu, \\
\displaystyle
\frac{1}{\eta^{\nu}}{n \choose \nu} \left( 1 - \frac{1}{\eta}
\right)^{n-\nu},
& n \ge \nu. 
\end{array}
\right.
\end{equation}
The inversion formula has a remarkable property that $\rho_{\nu}$
depends only on the ``tail'' of the photocount statistics for
$n \ge \nu$.
It has been shown that the inverse transformation can be applied to
experimentally determined photocount statistics for an arbitrary state
of the field, provided that the detection efficiency is higher that
50\% \cite{KissHerzPRA95}.

We will now discuss statistical properties of the photon number
distribution determined by photon counting within the general framework
developed in Sec.~\ref{Sec:Analysis}. The case of perfect detection is
trivial for statistical analysis. Therefore we will consider nonunit
detection efficiency, which is numerically compensated 
in the reconstruction process using the inverse Bernoulli
transformation according to Eq.~(\ref{Eq:InvBernoulli}). For all
examples presented here, the efficiency is $\eta=80\%$, which is well
above the $50\%$ stability limit.

\begin{figure}

\vspace{0.5cm}

\centerline{\epsfxsize=3in\epsffile{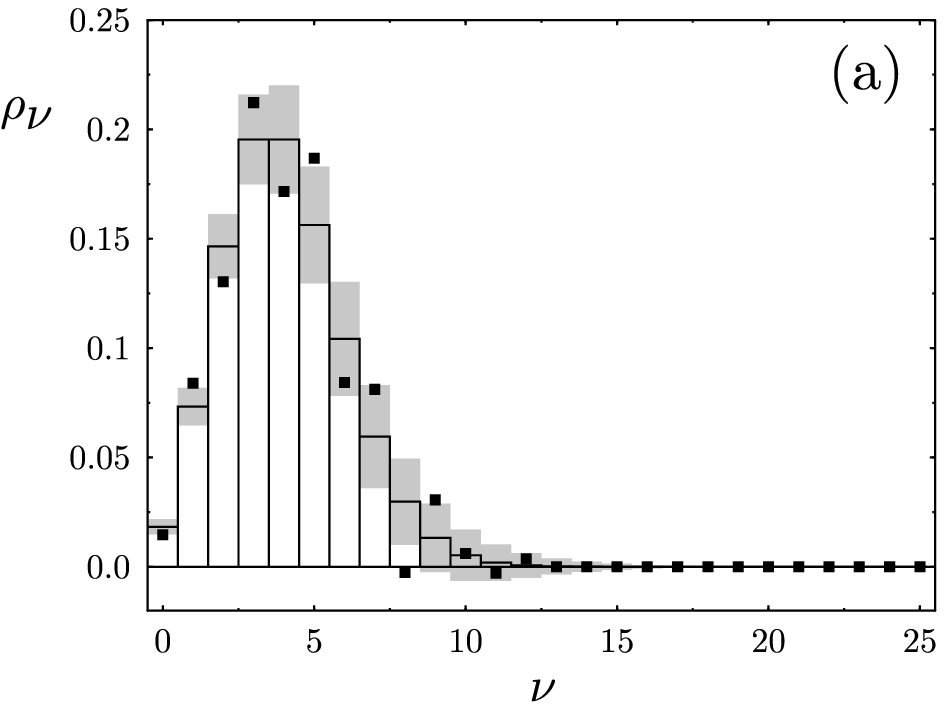}}

\vspace{0.5cm}

\centerline{\epsfxsize=3in\epsffile{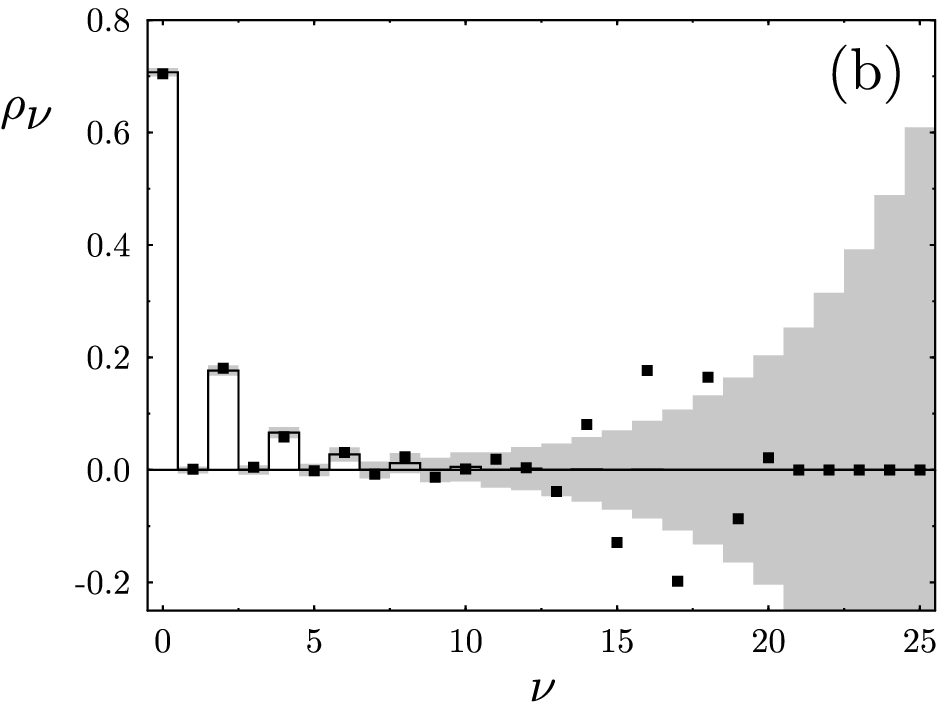}}

\vspace{0.5cm}

\centerline{\epsfxsize=3in\epsffile{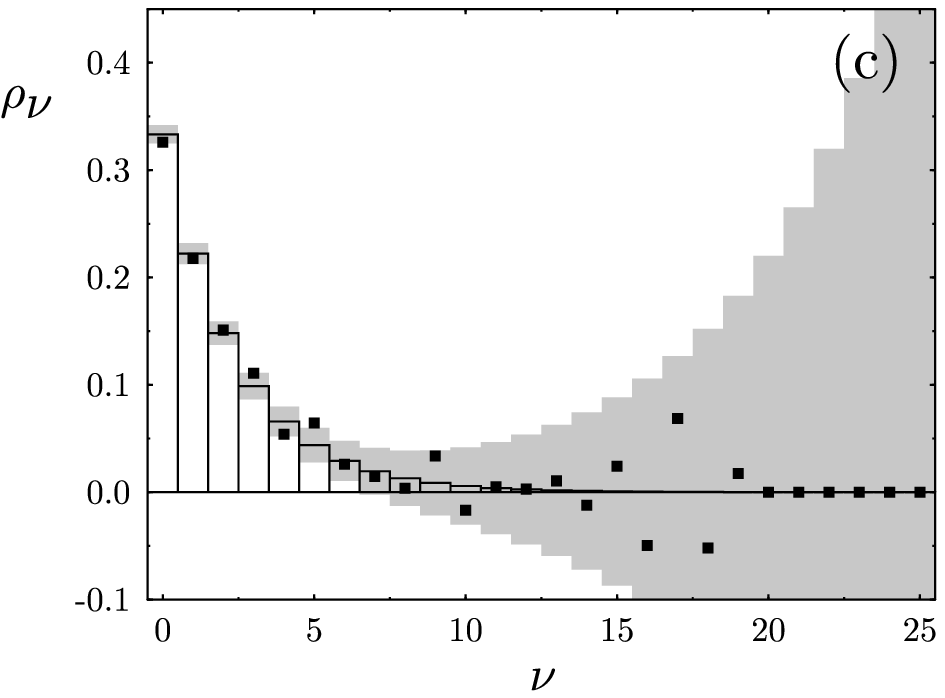}}

\vspace{0.5cm}

\caption{Reconstruction of the photon number distribution from
photon counting for (a) a coherent state with
$\langle\hat{n}\rangle=4$, (b) a squeezed vacuum state with
$\langle\hat{n}\rangle=1$, and (c) a thermal state with
$\langle\hat{n}\rangle=2$, from $N=4000$ runs in each case.
Monte Carlo simulations of a photon counting experiment, depicted with
points, are compared with exact values (solid lines), with the
statistical errors $[\text{Var}(\rho_{\nu})]^{1/2}$ marked as grey
areas. The detection efficiency is $\eta=80\%$.} 
\label{Fig:PhotonDistributions}
\end{figure}

\begin{figure}

\centerline{\epsffile{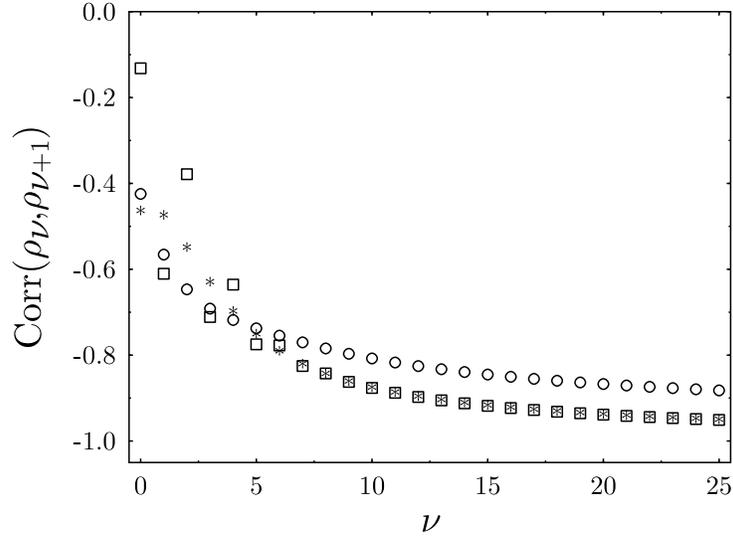}}

\vspace{0.5cm}

\caption{The correlation coefficient between the consecutive density
matrix elements $\text{Corr}(\rho_{\nu},\rho_{\nu+1})$, depicted for the
coherent state ($\circ$), the squeezed state ($\Box$), and the
thermal state ($\ast$) from Fig.~\protect\ref{Fig:PhotonDistributions}.}
\label{Fig:Correlations}
\end{figure}

In Fig.~\ref{Fig:PhotonDistributions} we depict the reconstructed
photon number distributions for a coherent state, a thermal state, and
a squeezed vacuum state. The mean
values $\text{E}(\rho_\nu)$ along with their
statistical errors $[\text{Var}(\rho_\nu)]^{1/2}$ are compared with
Monte Carlo realizations of a photon counting experiment, with
the number of runs $N=4000$. It is seen that for a thermal state and a
squeezed vacuum state, the statistical error 
of the probabilities $\rho_\nu$
grows unlimitedly with the
photon number $\nu$. However, any experimental histogram 
obtained from a finite number of runs ends up
for a certain count number, and therefore the reconstructed photon
statistics is zero above this number. An important feature that is
evidently seen in the Monte Carlo simulations, are correlations between
the consecutive matrix elements. The reconstructed photon number
distribution clearly exhibits oscillations around the true values.
This property can be quantified using the correlation
coefficient defined in Eq.~(\ref{Eq:Corr}), which we plot for all three
states in Fig.~\ref{Fig:Correlations}. For large $\nu$'s,
$\text{Corr}(\rho_\nu,\rho_{\nu+1})$ is close to its minimum allowed
value $-1$, which acknowledges that statistical correlations are indeed
significant.

These correlations affect any quantity computed from the reconstructed
photon number distribution. The parity operator is here a good example:
since in Eq.~(\ref{Eq:Parity}) we sum up consecutive $\rho_{\nu}$'s with
opposite signs, their statistical deviations do not add randomly,
but rather contribute with the same sign. Consequently, the statistical
error of the evaluated parity operator may be huge. It is therefore
interesting to study this case in detail. In principle, we could 
obtain statistical properties of the reconstructed
parity using the covariance matrix for the photon distribution. 
However, it will be more instructive to express
the parity directly in terms of the photocount statistics, and then to
apply the statistical analysis to this reconstruction recipe. 
This route is completely
equivalent to studying evaluation of the parity
 via the photon number distribution, as all
transformations of the experimental data, which we consider here, are
linear. 

A simple calculation
combining Eqs.~(\ref{Eq:ParityOp}) and (\ref{Eq:InvBernoulli}) shows that:
\begin{equation}
\label{Eq:Paritypn}
\hat{\Pi} = \sum_{n=0}^{K} \left( 1 - \frac{2}{\eta} \right)^n 
\hat{p}_n,
\end{equation}
where we have introduced in the upper summation limit 
a cut-off parameter $K$ for the photocount
number.  This formula clearly demonstrates pathologies related to the
determination of the parity operator.
For any $\eta<1$, the factor $(1-2/\eta)^n$ is not bounded,
which makes the convergence of the whole series questionable 
in the limit $K\rightarrow\infty$.
Of course, for an experimental histogram the summation is
always finite, but the exploding factor amplifies contribution from
the ``tail'' of the histogram, where usually only few events are
recorded, and consequently statistical errors are significant. 

\begin{figure}

\noindent\epsfxsize=3in\epsffile{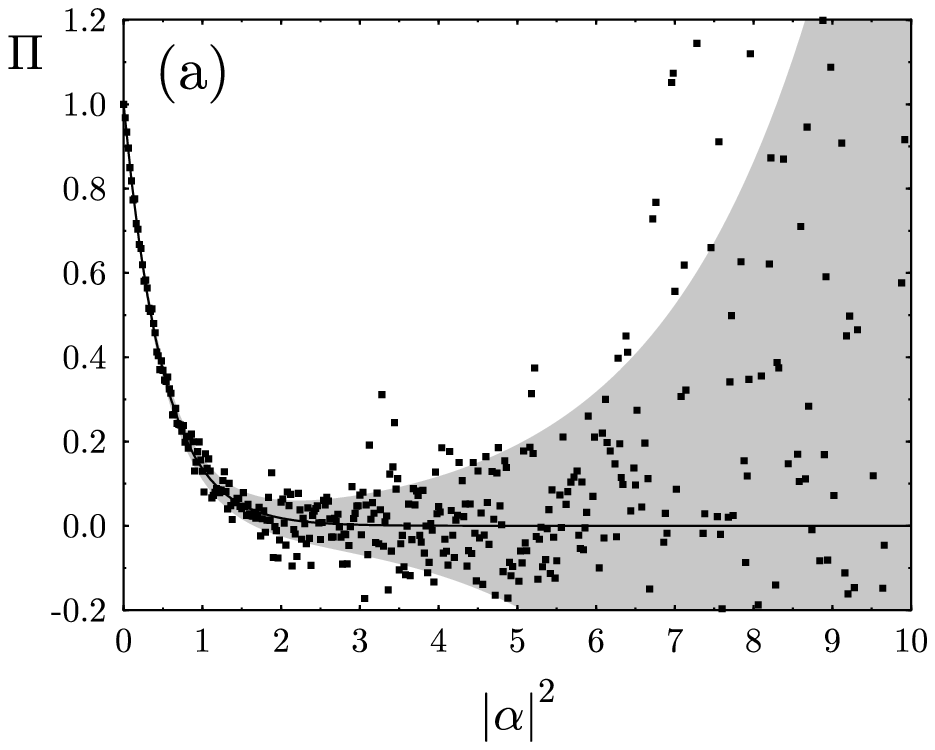}
\epsfxsize=3in\epsffile{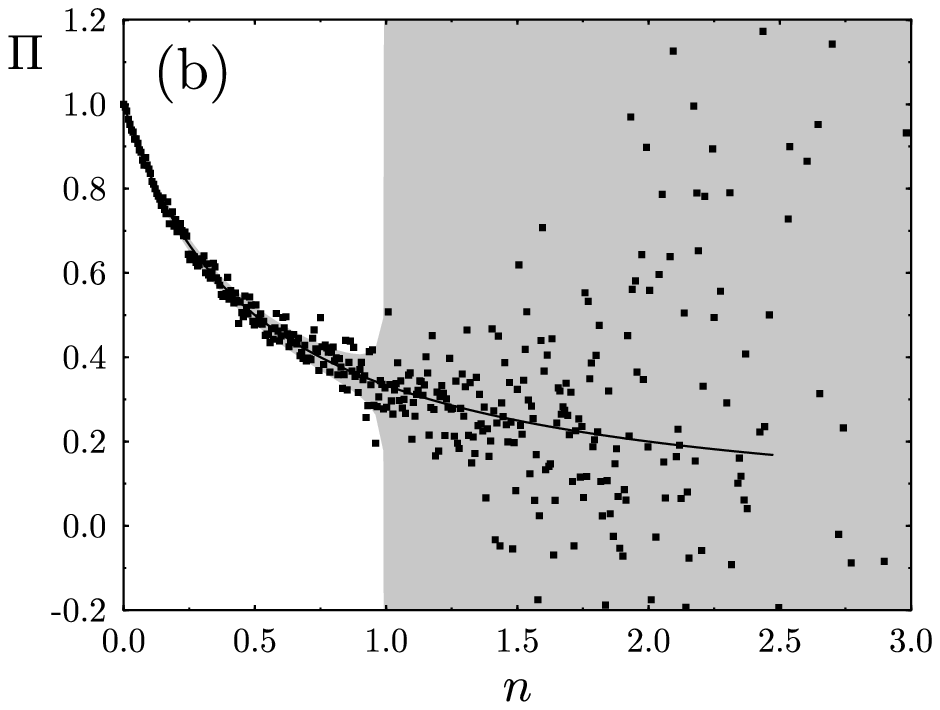}

\caption{Determination of the parity operator for (a) coherent states
and (b) thermal states with the increasing average photon number. Each
square represents the parity evaluated from Monte Carlo simulated
photon statistics with $N=4000$ runs
for a given average photon number. The solid lines and the grey areas
depict the mean value $\text{E}(\Pi)$ and
the error $[\text{Var}(\Pi)]^{1/2}$.}
\label{Fig:ParityCounting}
\end{figure}

Let us study these pathologies more closely using examples of a
coherent state and a thermal state. For a coherent state
$|\alpha\rangle$, both the expressions for $\text{E}(\Pi)$ and
$\text{Var}(\Pi)$ are convergent with $K\rightarrow\infty$. However,
the variance, given by the formula:
\begin{equation}
\text{Var}(\Pi^{\text{coh}}) = \frac{1}{N} \left[ \exp \left(
\frac{4(1-\eta)}{\eta} |\alpha|^2 \right) - \exp(-4|\alpha|^2)
\right],
\end{equation}
grows very rapidly with the coherent state amplitude $\alpha$, when
the number of runs $N$ is fixed.
For a thermal state with the average photon number $\bar{n}$, the
matter becomes more delicate. When $K\rightarrow\infty$,
the series (\ref{Eq:Paritypn}) is
convergent only for $\bar{n} < 1/[2(1-\eta)]$, which for $\eta=80\%$
gives just $2.5$ photons. Even when the mean value exists, the
variance is finite only for $\bar{n}<\eta/[4(1-\eta)]$ and equals:
\begin{equation}
\text{Var}(\Pi^{\text{th}}) = 
\frac{1}{N}\left( \frac{\eta}{\eta - 4 \bar{n}(1-\eta)}
- \frac{1}{(1 + 2 \bar{n})^2} \right).
\end{equation}
We illustrate these results with Fig.~\ref{Fig:ParityCounting},
depicting Monte Carlo simulations for various average photon
numbers. For coherent states, statistical fluctuations can in
principle be suppressed by increasing the number of runs. For thermal
states, the situation is worse: when $\bar{n}\ge 1$, the variance
cannot even be used as a measure of statistical uncertainty. 

Let us recall that the bound $\eta > 50\%$ for the stability of the
inverse Bernoulli transformation is independent of the state to be
measured. It has been obtained from the requirement that in the limit
$K\rightarrow\infty$ both $\text{E}(\rho_\nu)$ and
$\text{Var}(\rho_\nu)$ should converge. The example with the parity
operator clearly shows, that the condition $\eta>50\%$ does not
guarantee that the reconstructed photon number distribution can be
safely used to determine an arbitrary well--behaved phase
independent observable. Thus, imperfect detection is inevitably
connected with some loss of the information on the measured quantum
state.

We have noted that as long as finite, experimental data are concerned,
evaluation of observables via intermediate
quantities is equivalent to expressing them directly in terms of the
measured probability distributions. 
One might try to circumvent the $\eta>50\%$ bound for
reconstructing
the photon statistics by applying the inverse Bernoulli transformation
in two or more steps,
and compensating in each step only a fraction of its inefficiency. Of
course, such a strategy must fail, as for any finite sample of
experimental data this treatment is equivalent to a single
transformation which is unstable. In many--step processing this
instability would be reflected in increasing correlations and
statistical errors exploding to infinity. 

\subsection{Random phase homodyne detection}

Random phase homodyne detection is a recently developed technique for
measuring phase--independent properties of optical radiation, which
goes beyond certain limitations of plain photon counting
\cite{MunrBoggPRA95}. Data recorded in this scheme is the difference of
counts on two photodetectors measuring superposition of the signal
field with a strong coherent local oscillator. The count difference is
rescaled by the local oscillator amplitude, and the resulting
stochastic variable $x$ can be treated with a good approximation as a
continuous one. The photon number distribution is reconstructed from
the random phase homodyne statistics $p(x)$ by integrating it with pattern
functions $f_{\nu}(x)$ \cite{PatternFunctions}:
\begin{equation}
\langle\hat{\rho}_{\nu}\rangle
 = \int_{-\infty}^{\infty} \text{d}x \, f_{\nu}(x)
p(x).
\end{equation}
A convenient method for numerical evaluation of the pattern functions
has been described in Ref.~\cite{LeonMunrOC96}. 

Let us now discuss statistical properties of the homodyne scheme in
its discretized version used in experiments, when the rescaled count
difference is divided into finite width bins. 
As the local oscillator phase is
random, the setup has no controllable parameters, and the statistics
of the observables is fully determined by $p(x)$.
Statistical errors of the density matrix in the Fock basis
reconstructed via homodyne detection have been studied in
Ref.~\cite{DAriMaccQSO97}.
Here we will focus our attention on statistical
correlations exhibited by the diagonal 
density matrix elements, and their further
utilization for evaluating phase--independent observables. 

\begin{figure}

\centerline{\epsfxsize=3in\epsffile{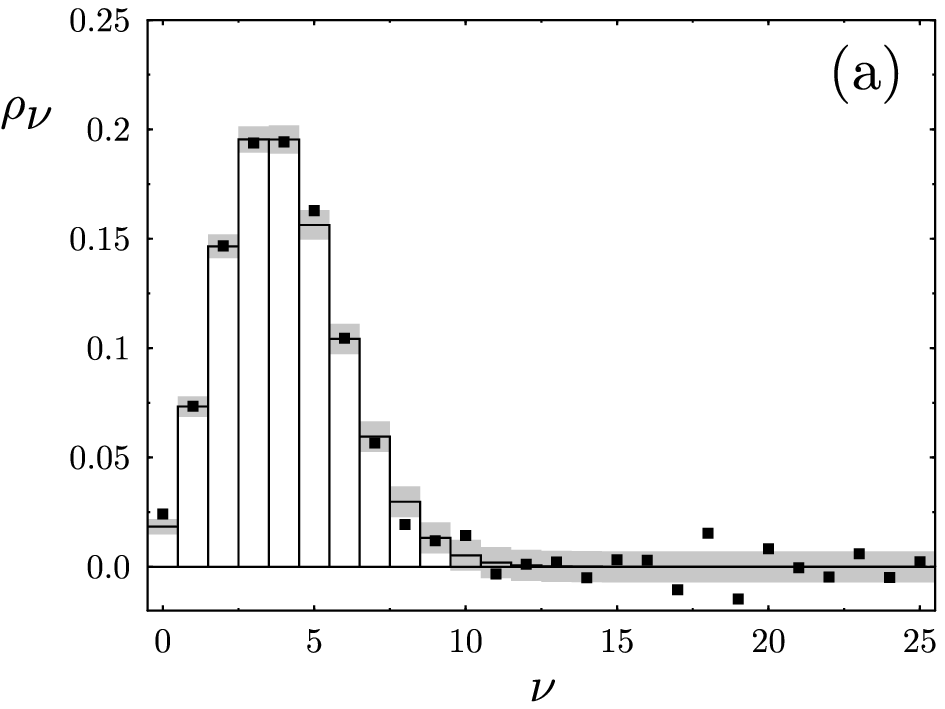}}

\vspace{0.5cm}

\centerline{\epsfxsize=3in\epsffile{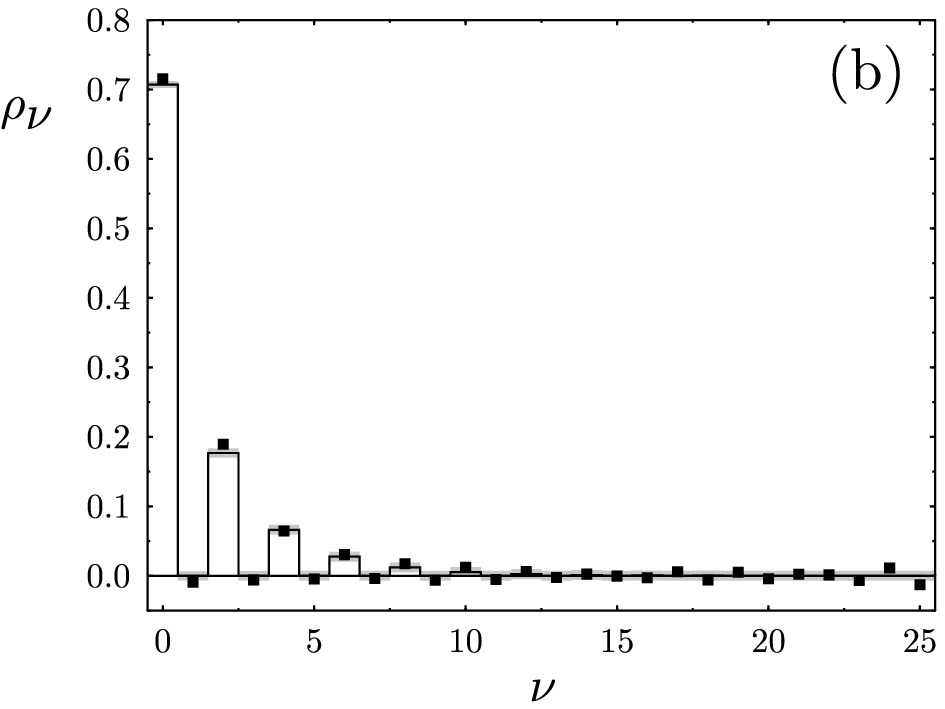}}

\vspace{0.5cm}

\centerline{\epsfxsize=3in\epsffile{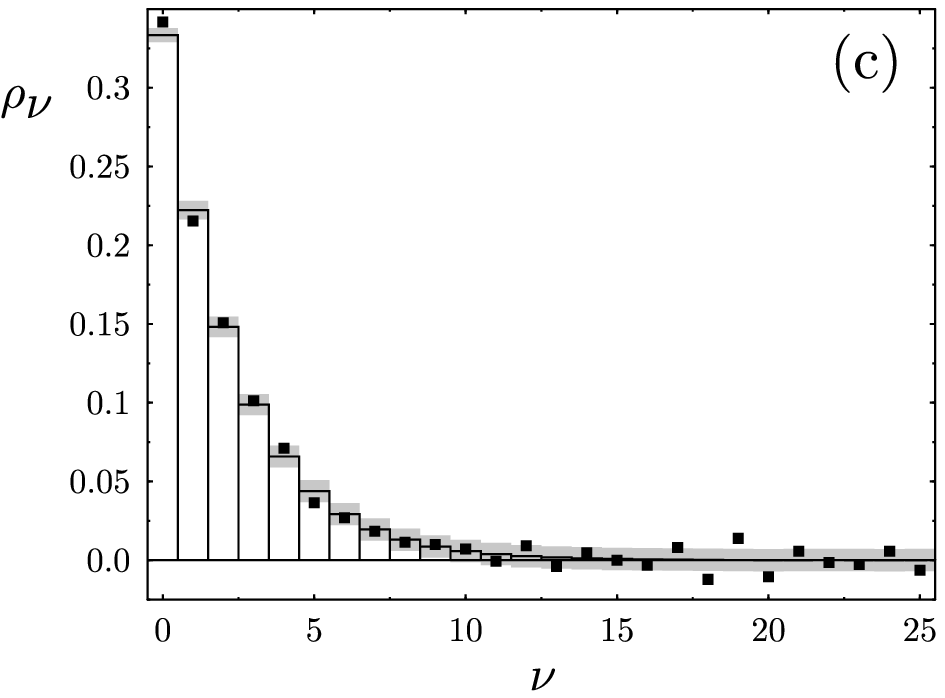}}

\vspace{0.5cm}

\caption{Random phase homodyne reconstruction of the
photon number distribution for (a) the coherent state (b)
the squeezed state and (c) the thermal state, all states with the
same photon numbers as in Fig.~\protect\ref{Fig:PhotonDistributions}.
The range of the homodyne variable is restricted to the interval $-6\le x
\le 6$ divided into 1200 bins. The simulated homodyne statistics is obtained
from $N=4 \cdot 10^4$ Monte Carlo events.}
\label{Fig:HomoPhotonDist}
\end{figure}

\begin{figure}

\centerline{\epsffile{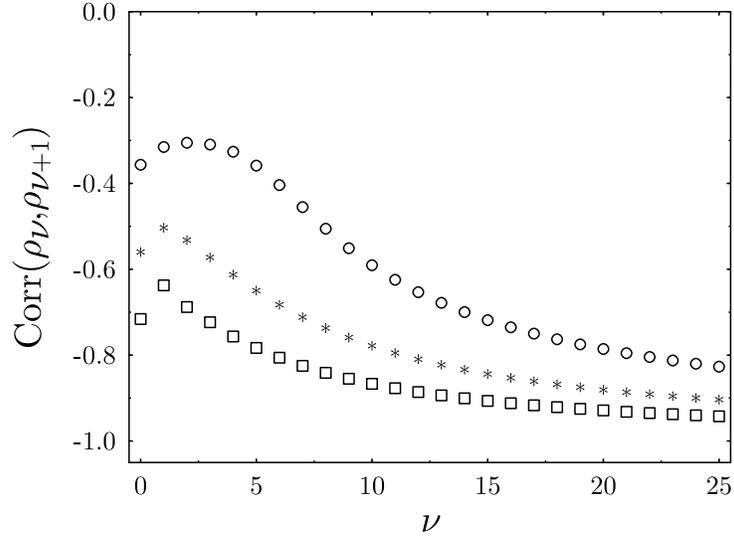}}

\vspace{0.5cm}

\caption{The correlation coefficient for the consecutive photon
number probabilities for the coherent state ($\circ$), the squeezed
state ($\Box$) and the thermal state ($\ast$).}
\label{Fig:HomoCorr}
\end{figure}

We will consider the unit detection efficiency $\eta=1$, with no
compensation in the processing of the experimental data. 
This is the most regular case from the numerical point of view. When
$\eta < 1$ and the compensation is employed, the statistical errors are
known to increase dramatically \cite{DAriMaccQSO97}.
In Fig.~\ref{Fig:HomoPhotonDist} we depict 
the homodyne reconstruction of the photon
number distribution for the three states discussed in the previous
subsection. For large $\nu$, the statistical errors tend to a fixed
value $\sqrt{2/N}$, which has been explained by D'Ariano {\em et al.}
using the asymptotic form of the pattern functions \cite{DAriMaccQSO97}.
Again, Monte Carlo simulations suggest that the reconstructed density
matrix elements are correlated, which is confirmed by the correlation
coefficient for the consecutive photon number probabilities, plotted in
Fig.~\ref{Fig:HomoCorr}. A simple 
analytical calculation involving the asymptotic
form of the pattern functions shows, that for large $\nu$ this
coefficient tends to its minimum value $-1$. 

\begin{figure}

\centerline{\epsffile{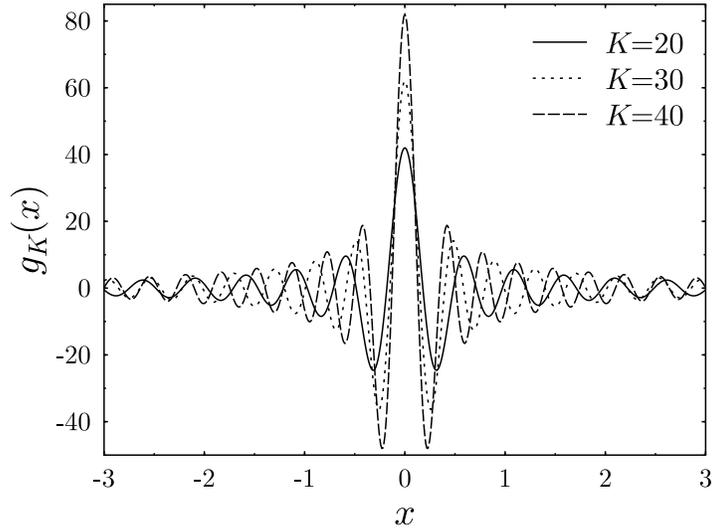}}

\vspace{0.5cm}

\caption{Regularized kernel functions for the parity operator
$g_K(x)$
evaluated as a finite sum of the Fock states pattern functions,
for increasing values of the cut--off parameter $K$.}
\label{Fig:ParityPatterns}
\end{figure}

\begin{figure}

\centerline{\epsffile{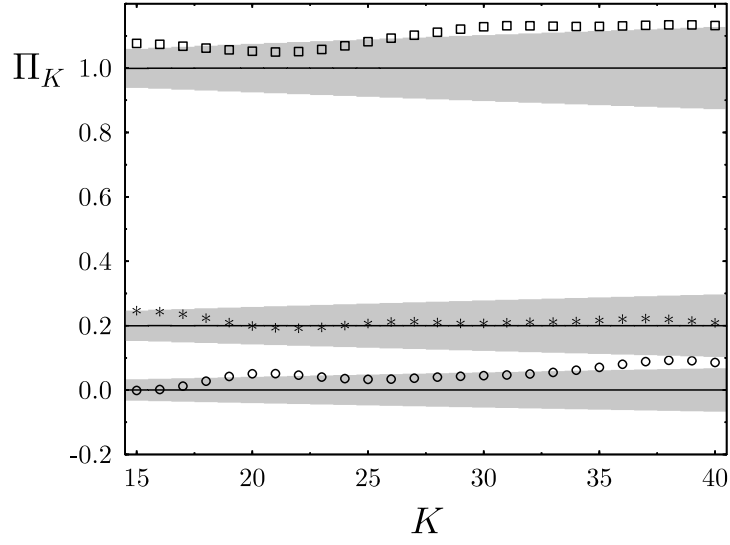}}

\vspace{0.5cm}

\caption{Reconstruction of the truncated parity operator 
$\hat{\Pi}_K$ for the coherent
state ($\circ$), the squeezed state ($\Box$), and the thermal state
($\ast$) with various values of the cut--off parameter $K$,
using the same Monte Carlo homodyne statistics 
as in Fig.~\protect\ref{Fig:HomoPhotonDist}. The simulations are compared
with the corresponding mean values $\text{E}(\Pi_K)$ and errors
$[\text{Var}(\Pi_K)]^{1/2}$ plotted as solid lines surrounded by grey
areas.}
\label{Fig:HomoParity}
\end{figure}

One may now expect that no subtleties can be hidden in using the
reconstructed photon number distribution to evaluate the parity operator
according to Eq.~(\ref{Eq:Parity}).  However, let us recall that the
parity operator is equal, up to a 
multiplicative constant, to the Wigner function at
the phase space origin \cite{RoyePRA77,MoyaKnigPRA93}. 
The Wigner function is
related to the homodyne statistics via the inverse Radon transformation,
which is singular. In particular, applying this transformation
for the phase space point $(0,0)$ we obtain the following expression
for the parity operator in terms of the homodyne statistics
\cite{LeonJexPRA94}:
\begin{equation}
\langle\hat{\Pi}\rangle
 = \frac{1}{2} \int_{-\infty}^{\infty} \text{d}x \, p(x)
\frac{\text{d}}{\text{d}x} P\frac{1}{x},
\end{equation}
where $P$ denotes the principal value. Due to the singularity of the
inverse Radon transform, its application to experimental data has to
be preceded by a special filtering procedure. This feature must
somehow show up, when we evaluate the parity
operator from the reconstructed photon statistics. In order to analyse
this problem in detail let us discuss evaluation of the 
truncated parity operator $\hat{\Pi}_K$
from a finite part of the photon number distribution:
\begin{equation}
\langle\hat{\Pi}_{K}\rangle
= \sum_{\nu=0}^{K} (-1)^\nu \langle\hat{\rho}_{\nu}\rangle
= \int_{-\infty}^{\infty} g_K(x) p(x),
\end{equation}
where 
\begin{equation}
g_{K}(x) = \sum_{\nu=0}^{K} (-1)^{\nu} f_{\nu}(x)
\end{equation}
can be considered to be a regularized kernel function for the parity
operator. In Fig.~\ref{Fig:ParityPatterns}
we plot this function for increasing
values of the cut--off parameter $K$. It is seen that the singularity
of the kernel function in the limit $K\rightarrow\infty$ is reflected
by an oscillatory behaviour around $x=0$ with growing both the
amplitude and the frequency. 
This amplifies the statistical uncertainty of the
experimental homodyne data. 
In Fig.~\ref{Fig:HomoParity} we show determination
of the parity operator for the three states discussed before, using
increasing values of the cut--off parameter $K$. Though we are in the
region where the true photon number distribution is negligibly small,
addition of subsequent matrix elements increases the statistical
error in an approximately linear manner. This is easily understood,
if we look again at the reconstructed photon number distributions:
increasing $K$ by one means a contribution of the order of
$\sqrt{2/N}$ added to the statistical uncertainty, and, moreover,
these contributions tend to have the same sign due to correlations
between the consecutive matrix elements. Thus, determination of the
parity operator from homodyne statistics requires an application of a
certain regularization procedure. It may be either the filtering used in
tomographic back--projection algorithms,
or the cut--off of the photon number
distribution. The statistical uncertainty of the final outcome is
eventually a result of an interplay between the number of experimental
runs and the applied regularization scheme. 

Finally, let us briefly comment on the compensation for the nonunit
efficiency of the homodyne detector. First, one might think of applying
a two--mode inverse Bernoulli transformation directly to the joint count
statistics on the detectors. However, it is impossible in the homodyne
scheme to resolve contributions from single absorbed photons due to
high intensity of the detected fields.  The inverse Bernoulli
transformation has no continuous limit, as consecutive count
probabilities are added with opposite signs.
Nevertheless, the
nonunit detection efficiency can be taken into account in the pattern
functions \cite{PatternFunctions}. 
In this case the statistical errors increase dramatically,
and explode with $\nu\rightarrow\infty$, which makes determination of
the parity operator even more problematic. This is easily understood
within the phase space picture: the distributions measured by an
imperfect homodyne detector are smeared--out by a convolution with a
Gaussian function 
\cite{VogeGrabPRA93,LeonPaulPRA93}. Evaluation of the parity
operator, or equivalently, the Wigner function at the phase space
origin requires application of a deconvolution procedure, which
enormously amplifies the statistical error \cite{LeonPaulJMO94}. 

\section{Conclusions}
\label{Sec:Conclusions}
We have presented a complete statistical analysis of determining
quantum observables in optical measurement schemes based on
photodetection. We have derived an exact expression for the generating
function characterizing statistical moments of the reconstructed
observables, which, in particular, provides formulae for statistical errors
and correlations between the determined quantities. 
These general results have been applied to the detection of
phase--independent properties of a single light mode using two schemes:
direct photon counting, and random phase homodyne detection. This study
has revealed difficulties related to the completeness of the
reconstructed information on the quantum state: in some cases the
parity observable, which is a well--behaved bounded operator,
effectively cannot be evaluated from the reconstructed data due to
the exploding statistical error.

We have recalled that the parity operator is directly related to the
value of the Wigner function at the phase space origin.  Thus, our
example can also be interpreted as a particular case of the
transformation between two representations of the quantum state: in
fact, we have considered evaluation of the Wigner function at a
specific point $(0,0)$ from the relevant elements of the density matrix
in the Fock basis.  Therefore, our discussion exemplifies subtleties
related to the transition between various quantum state
representations, when we deal with data reconstructed in experiments.
Though a certain representation can be determined with the statistical
uncertainty which seems to be reasonably small, it effectively cannot
be converted to another one due to accumulating statistical errors.

The presented study suggests that the notion of completeness in quantum
state measurements should inherently take into account the statistical
uncertainty. From a theoretical point of view, the quantum state can be
characterized in many different ways which are equivalent as long as
expectation values of quantum operators are known with perfect
accuracy.  In a real experiment, however, we always have to keep in
mind the specific experimental scheme used to perform the measurement.
This scheme defines statistical properties of the reconstructed
quantities, and may effectively limit the available information on the
quantum state. Determination of a family of observables does not
automatically guarantee the feasibility of reconstructing the
expectation value of an arbitrary well behaved operator. Reconstruction
of any observable should be preceded by an analysis how significantly
the final result is affected by 
statistical noise corrupting the raw experimental data.

\section*{Acknowledgements}

The author is indebted to Professor K.~W\'{o}dkiewicz for numerous
discussions and valuable comments on the manuscript. This research was
supported by the Polish KBN grant No.\ 2P03B~002~14 and by 
Stypendium Krajowe dla M{\l}odych Naukowc\'{o}w Fundacji na rzecz
Nauki Polskiej.

\vspace{1cm}

\vspace{1cm}

\end{document}